\documentclass{vgtc}                          

\ifpdf
  \pdfoutput=1\relax                   
  \usepackage{graphicx}                
  \DeclareGraphicsExtensions{.pdf,.png,.jpg,.jpeg} 
\else
  \usepackage{graphicx}                
  \DeclareGraphicsExtensions{.eps}     
\fi%

\graphicspath{{figures/}{pictures/}{images/}{./}} 

\usepackage{microtype}                 
\PassOptionsToPackage{warn}{textcomp}  
\usepackage{textcomp}                  
\usepackage{mathptmx}                  
\usepackage{times}                     
\usepackage{cite}                      
\usepackage{tabu}                      
\usepackage{booktabs}                  

\onlineid{0}

\vgtccategory{Research}

\title{Visualizing Non-Fungible Token Ethics: A Case Study On CryptoPunks}

\author{Yufan Zhang\thanks{e-mail: yz605@duke.edu}\\ %
        \parbox{1.4in}{\scriptsize \centering Duke Kunshan University \\ SciEcon CIC} %
\and Zichao Chen\thanks{e-mail: zc142@duke.edu}\\ %
     \parbox{1.4in}{\scriptsize \centering Duke Kunshan University \\ SciEcon CIC} %
\and Luyao Zhang\thanks{e-mail: lz183@duke.edu}\\ %
     \parbox{1.4in}{\scriptsize \centering Duke Kunshan University \\ SciEcon CIC} %
\and Xin Tong\thanks{e-mail: xin.tong@dukekunshan.edu.cn}\\ %
     \scriptsize Duke Kunshan University %
}

\abstract{As a blockchain-based application, Non-Fungible Token (NFT) has received worldwide attention over the past few years. Digital artwork is the main form of NFT that can be stored on different blockchains. Although the NFT market is rapidly developing, we observed potential ethical and racial fairness issues in the design of NFT artworks due to a  lack of ethical guidelines or censorship. Therefore, we investigated CryptoPunks, the most famous collection in the NFT market, to explore and visualize its potential ethical issues. We explored the ethical issues from three aspects: design, trading transactions, and related topics on Twitter. We scraped data from Twitter and Dune Analytics using python libraries, Twitter crawler, and sentiment analysis tools. Our five visualizations implied that 1.6 times more male punks were created in the initial design process than the female ones. And the male ones have a higher average selling price than females; lighter-skinned punks tend to sell for higher prices. The results of our study and visualizations provide a preliminary exploration of CryptoPunks and further inspire future ethical-related investigation and research in the NFT domain.%
} 

\begin{document}


\firstsection{Introduction}

\maketitle
Non-fungible token (NFT) is a unit of data stored on the blockchain that certifies a digital asset to be unique \cite{tabora_2021_looking}. The NFT market achieved more than 2 billion dollars in transactions in the first quarter of 2021 — which is 2000\% higher than the fourth quarter of 2020 \cite{frank_2021_nft}. Besides, many famous artists and companies have joined the NFT market \cite{swant_as, farrington_why}. Numerous NFT digital avatar collections are created based on attributes like the avatar’s gender, skin color, and accessories. CryptoPunks \cite{cryptopunks} is one of the most popular NFT collections that offers more than 10,000 unique collectible character images, selling on every mainstream NFT marketplace. 

However, potential ethical problems arise in the fledgling NFT market without a regulatory mechanism and central government. For example, CryptoPunks with different genders and skin colors have significantly different price trends \cite{egkolfopoulou_2021_even}. Moreover, the algorithmic or manual generation processes of NFT usually remain unrevealed. Although CryptoPunks creators claimed that each artwork is generated algorithmically, the punks showed imbalance attributes in gender and race \cite{cryptopunks}.

Therefore, our research questions of this project are:

\begin{itemize}
\item What potential gender and racial ethical concerns exist in NFT artwork collections in the generation and trading process, such as CryptoPunks? 

\item What are Twitter users’ perceptions of NFT ethical problems?
\end{itemize}

\section{Methodology}

\subsection{Data Source}
The data are from two data sources: blockchain transaction records of CryptoPunks token from DuneAnalytics \cite{dune} ranging from the launch date of CryptoPunks, June 23, 2017, to May 7, 2022, and the Twitter threads with at least five likes from Twitter \cite{twitter_2000_twitter} with related topic keywords including “NFT,” “CryptoPunk(s),” “ethic,” “informed consent,” “transparency,” “accountability,” “privacy,” “fairness”, “trust,” “gender,” “ethnicity,” “skin tone,” and “skin color.”

\subsection{Data Processing}

\subsubsection{Construct databases}

Using the Python Pandas \cite{pandas_2018_python} and Numpy \cite{numpy_2009_numpy} libraries, we construct four databases: transaction database (16,823 transaction records), token database (10,000 punks), trader database (5,911 addresses), and tweet database (83,568 tweets), whose primary keys are the transaction ID, CryptoPunks ID, Ethereum address, Twitter thread ID respectively.

\subsubsection{Extract visualization data}

To depict the distribution of CryptoPunks with different attributes, we group the CryptoPunks tokens with their attributes. Specifically, the attributes of CryptoPunks are categorized into four levels: the type (\textit{e.g.}, human), the gender (\textit{e.g.}, male), the skin tone (\textit{e.g.}, dark), and the number of attributes (\textit{e.g.}, 3 attributes). Furthermore, the transaction database is merged with the token database to add the CryptoPunks ethics-related attributes (\textit{i.e.}, gender and skin tone) into each transaction record. Using the Python NetworkX \cite{networkx} library, we build the CryptoPunks transaction network for each year from 2017 to 2022 with the addresses as the nodes and the transaction as the edges. 

\subsubsection{Conduct sentiment analysis}

Using the Python flairNLP \cite{a2020_flairnlpflair} library, we conducted sentiment analysis on the tweet content in the tweet database. Unlike the rule-based sentiment analysis toolkit, such as Textblob \cite{a2018_textblob}, flairNLP processes the texts via the deep neural network, which generally yields better model performance in Natural Language Processing (NLP) tasks such as sentiment analysis. Specifically, all the tweets are classified into positive, neutral, or negative sentiments.

\subsection{Data Visualizations}

We construct five visualizations using the Python Plotly library \cite{modern}. First, we create an interactive Sankey diagram to illustrate the distribution of CryptoPunks with different attributes. Second, we apply the Time River design idiom to visualize the price change of different combinations of attributes. Third, we create a scatter plot to depict the price disparity of skin tone. Fourth, we create a circular network visualization based on the undirected graph of the transaction network by year. Last, we create the bar chart and word cloud figures using Python WordCloud \cite{wordcloud} package to visualize Twitter users’ sentiment towards NFT-related ethical topics. A detailed introduction to the visualization results will be shown in the results section.

\section{Results}


\subsection{CryptoPunks Generating Distribution}


\begin{figure}[h]
    \centering
    \includegraphics[width=0.4\textwidth]{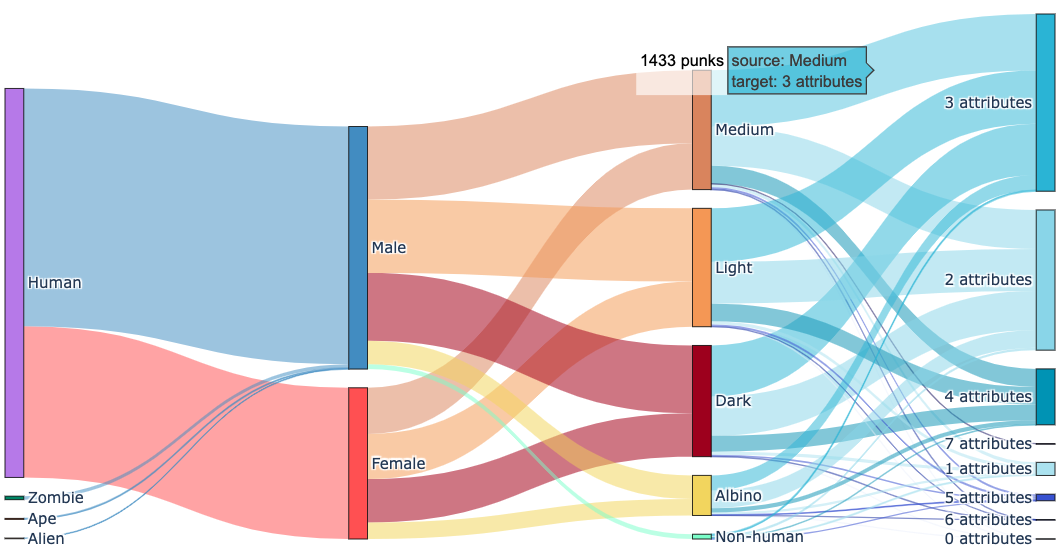}
    \caption{Sankey Diagram of the Attributes Distribution}
    \label{vis1}
\end{figure}

Knowing the initial amount of CryptoPunks provides insights of the design team's opinions of equality. Figure \ref{vis1} reveals intuitively that the number of genders in the original set of CryptoPunks is quite different. There are 6,039 male Punks, but only 3,840 female Punks. There were also differences in the number of punks with different skin colors, with dark skin having the lowest number.

\subsection{Price Trend of Punks in Different Skin Colors and Genders}

\begin{figure}[h]
    \centering
    \includegraphics[width=0.4\textwidth]{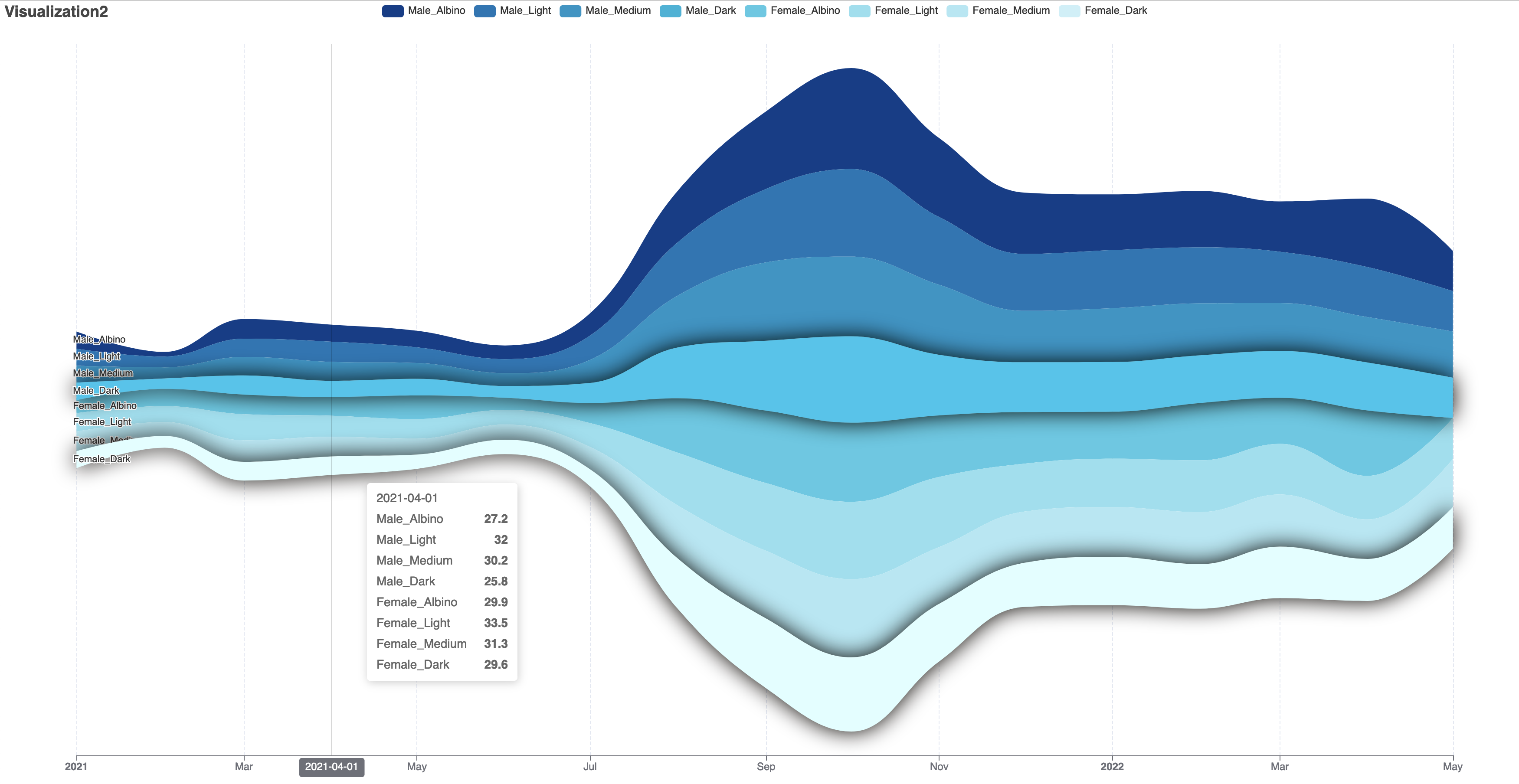}
    \caption{Time River Chart of the Price Trend}
    \label{vis2}
\end{figure}

In this section, we analyzed the price changes of eight types of Punks in the market and the differences between them, as shown in Figure \ref{vis2} (from top to bottom are  "Male\_Albino", "Male\_Light", "Male\_Medium", "Male\_Dark", "Female\_Albino", "Female\_Light", "Female\_Medium", "Female\_Dark"). Figure 2 indicates that the average transaction prices of male and light-skinned punks was higher than that of female and dark-skinned punks of the same period at most time from year 2017 to now. However, the price gaps among these groups were not very huge.

\subsection{Transaction Prices Sorted by Skin Color Type}

\begin{figure}[h]
    \centering
    \includegraphics[width=0.4\textwidth]{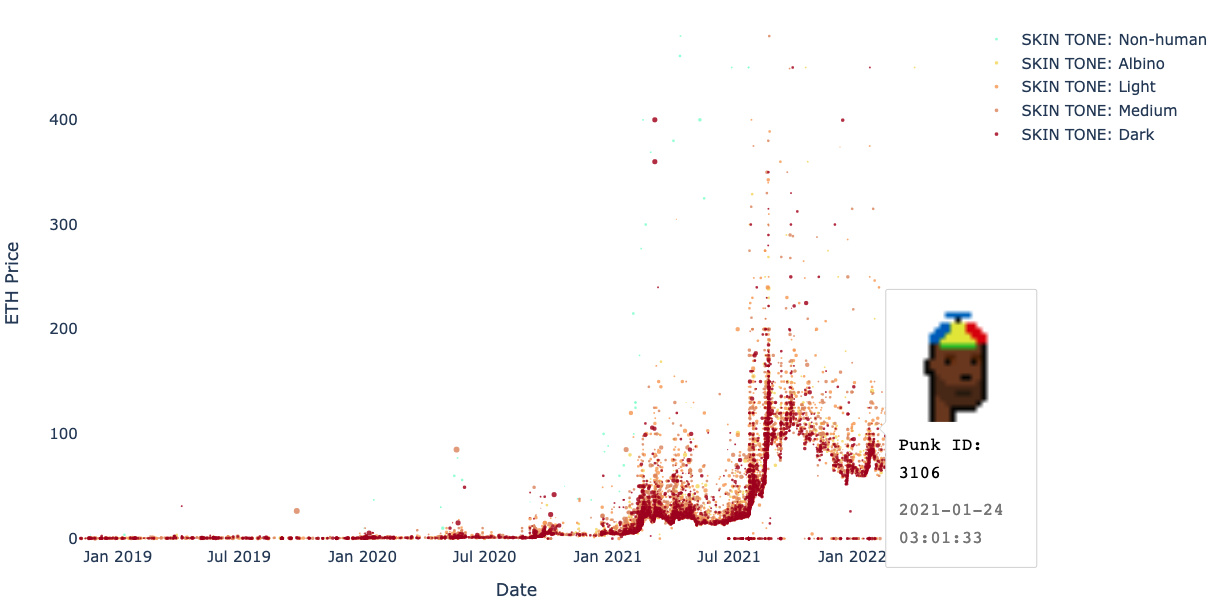}
    \caption{Scatter Plot of the Transaction Record}
    \label{vis3}
\end{figure}

The scatter plot visualizes the CryptoPunks transaction records and its skin tone. Each circle represents a transaction, and its color refers to the punk’s skin tone. Through interaction, users can see the CryptoPunks details and data for each transaction. In general, the dark spots were generally below the light ones, which implied that the price of dark Punk is generally lower than that of light Punk.

\subsection{CryptoPunks Transaction Network (by Year)}

\begin{figure}[h]
    \centering
    \includegraphics[width=0.4\textwidth]{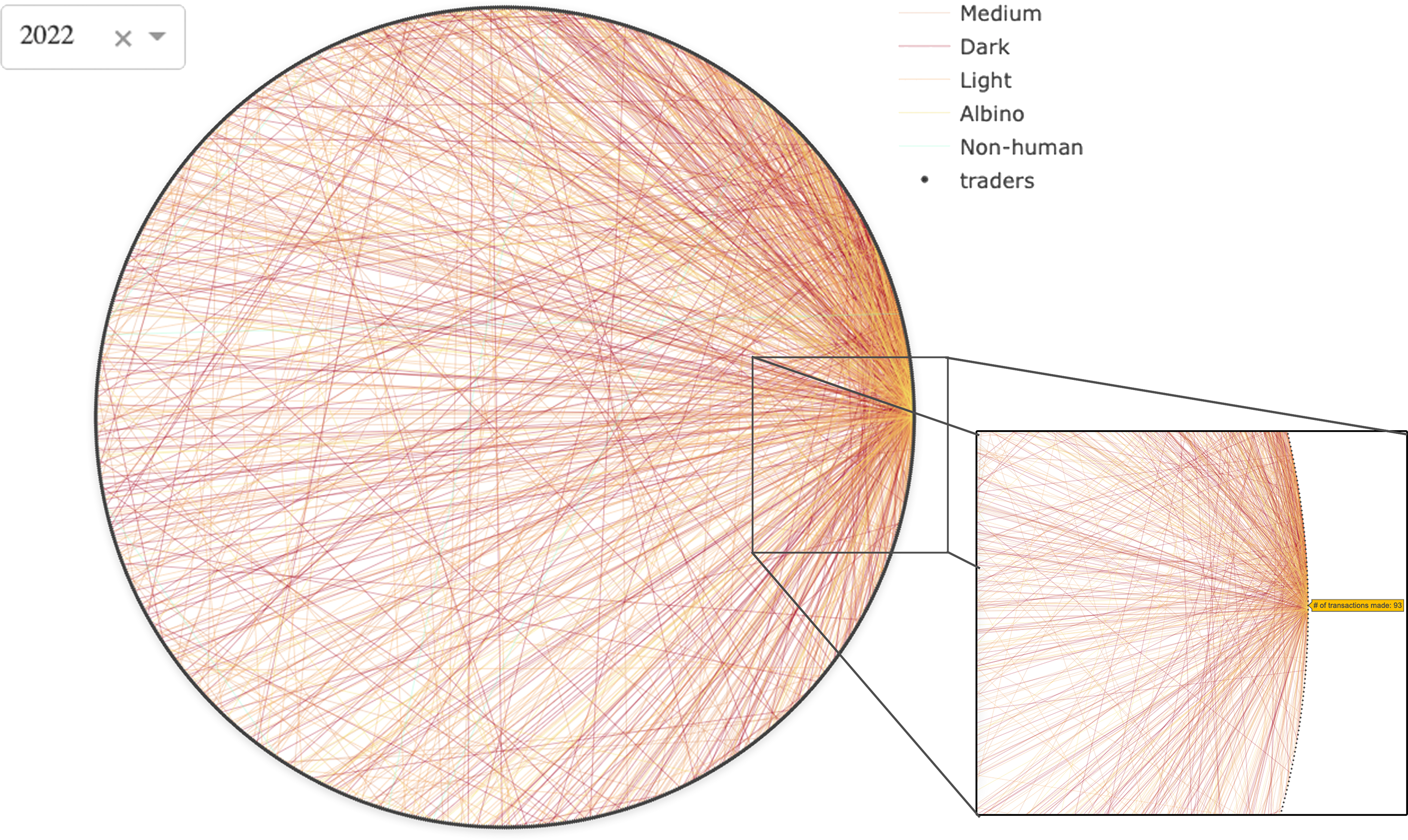}
    \caption{Circular Network Graph of the Transaction}
    \label{vis4}
\end{figure}

The circular network visualizes the transaction network of CryptoPunks by year, which explores CryptoPunks' current trading market and trading network. Each node refers to a trader who has participated in at least one transaction of CryptoPunks, and each edge refers to a transaction with the color representing the punk skin tone of the transaction. In this graph, the right side of the transaction line is the most dense, which means that prices in the CryptoPunks market are likely to be dominated by a few big business merchants or high frequency collectors.

\subsection{Twitter Analysis with Bar Chart \& Word Cloud}

\begin{figure}[h]
    \centering
    \includegraphics[width=0.4\textwidth]{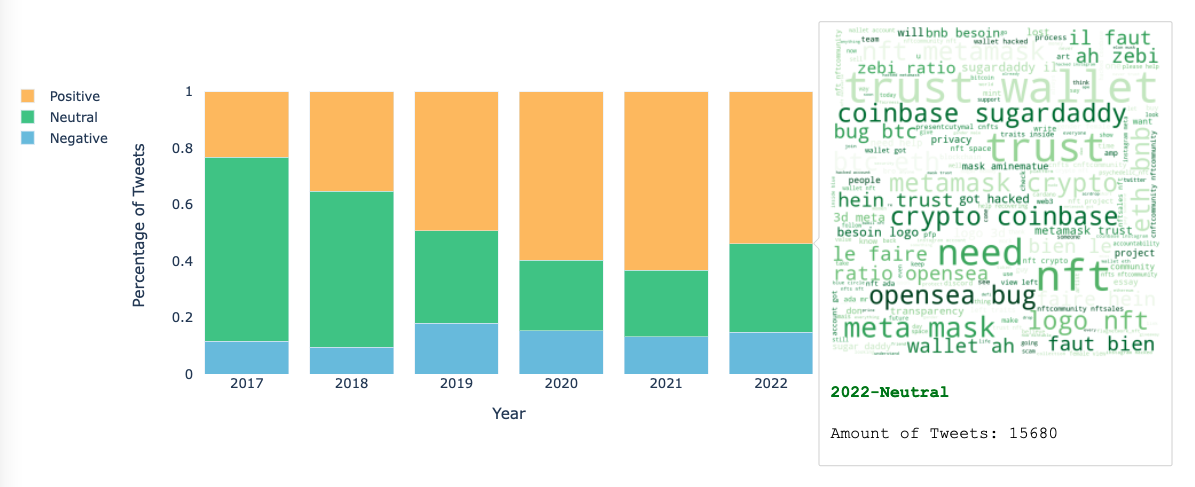}
    \caption{Bar Chart of the Word Cloud and Sentiment Analysis}
    \label{vis5}
\end{figure}

The bar chart and the word cloud visualizations in Figure \ref{vis5} reflected the sentiments of Twitter users' attitudes about NFT ethics. We did not identify many participants who are aware of gender and ethical inequities in NFT, which means that the ethical training and management of the NFT market is urgently needed and necessary.

\section{Discussions and Conclusion}

NFT is a scorching and popular market where the numbers of players are growing rapidly. In our project, we successfully collected and visualized data about the market transactions of CryptoPunks, an NFT artwork collection and tweets. Our five interactive visualizations demonstrated that different levels of gender and racial inequality have emerged both during the creation and trading processes of CryptoPunks. Our project, as a pioneer in this field, will resonate with the community and more players and inspire more researchers to study the ethical issues of blockchain.

\bibliographystyle{abbrv-doi}

\bibliography{chinavis}
\end{document}